**Abstract.** We compute the temperature profiles of accretion discs around rapidly rotating strange stars, using constant gravitational mass equilibrium sequences of these objects, considering the full effect of general relativity. Beyond a certain critical value of stellar angular momentum ($J$), we observe the radius ($r_{\mathrm{orb}}$) of the innermost stable circular orbit (ISCO) to increase with J (a property seen neither in rotating black holes nor in rotating neutron stars). The reason for this is traced to the crucial dependence of $dr_{\mathrm{orb}}/dJ$ on the rate of change of the radial gradient of the Keplerian angular velocity at $r_{\mathrm{orb}}$ with respect to $J$. The structure parameters and temperature profiles obtained are compared with those of neutron stars, as an attempt to provide signatures for distinguishing between the two. We show that when the full gamut of strange star equation of state models, with varying degrees of stiffness are considered, there exists a substantial overlap in properties of both neutron stars and strange stars. However, applying accretion disc model constraints to rule out stiff strange star equation of state models, we notice that neutron stars and strange stars exclusively occupy certain parameter spaces. This result implies the possibility of distinguishing these objects from each other by sensitive observations through future X–ray detectors.

# Temperature profiles of accretion discs around rapidly rotating strange stars in general relativity: a comparison with neutron stars

**Sudip Bhattacharyya[1,2], Arun V. Thampan[3], and Ignazio Bombaci[4]**

[1] Joint Astronomy Program, Indian Institute of Science, Bangalore 560012, INDIA

[2] Indian Institute of Astrophysics, Bangalore 560 034, INDIA
(sudip@physics.iisc.ernet.in; sudip@iiap.ernet.in)

[3] Inter-University Centre for Astronomy and Astrophysics (IUCAA), Pune 411 007, INDIA
(arun@iucaa.ernet.in)

[4] Dipartimento di Fisica "E. Fermi" Universitá di Pisa, and INFN Sezione di Pisa, via Buonarroti 2, 56127 Pisa, ITALY
(bombaci@df.unipi.it)

## 1. Introduction

Low mass X–ray binaries (LMXBs) are believed to contain either neutron stars (NSs) or black holes accreting from an evolved or main sequence dwarf companion that fills its Roche–lobe. The proximity of the companion in these systems cause matter to spiral in, forming an accretion disc around the central accretor. Observations of LMXBs can provide vital clues of the structure parameters of the accretors and, in particular for NSs, this can lead to constraining the property of the high density matter composing their interiors. Therefore, the estimation of the radius of the central accretor in SAX J1808.4 and 4U 1728-34 (Li et al. 1999a; Li et al. 1999b; Burderi & King 1998; Psaltis & Chakrabarty 1999) indicating the object to be more compact than stars composed of high density nuclear matter, acquires significance. These results moot alternate suggestion about the nature of the central accretors in at least some of the LMXBs.

In this regard, the *strange matter hypothesis*, formulated by Bodmer (1971) and Witten (1984) (see also Itoh 1970; Terazawa 1979), has received much attention recently. The hypothesis suggests strange quark matter (SQM, made up of u, d and s quarks), in equilibrium with weak interactions, to be the actual ground state of strongly interacting

---

matter rather than $^{56}$Fe. If this were true, under appropriate conditions, a phase transition within a NS (e.g. Olinto 1987; Cheng & Dai 1996; Bombaci & Datta 2000) could convert the entire system instantaneously into a conglomeration of strange matter or, as is commonly referred to in literature, strange stars (SSs).

It is of fundamental interest - both for particle physics and astrophysics - to know whether strange quark matter exists. Answering this question requires the ability to distinguish between SSs and NSs, both observationally as well as theoretically and this has been the motivation of several recent calculations (Xu et al. 2001; Gondek-Rosinska et al. 2000; Bombaci et al 2000; Zdunik 2000; Zdunik et al 2000a; Zdunik et al 2000b; Datta et al 2000; Stergioulas et al 1999; Gourgoulhon et al 1999; Xu et al 1999; Gondek & Zdunik 1999; Bulik et al. 1999; Lu 1998; Madsen 1998). One of the most basic difference between SSs and NSs is the mass–radius relationship (Alcock, Farhi & Olinto 1986): while for NSs, this is an inverse relationship (radius decreasing for increasing mass), for SSs there exist a positive relationship (radius increases with increasing mass) [1]. In addition to this difference, due to SSs being self–bound objects, there also exists the possibility of having configurations with arbitrarily small masses; NSs on the other hand, have a minimum allowed mass (e.g. Shapiro & Teukolsky 1983; Glendenning 1995; and more recently, Gondek et al. 1997; Gondek et al. 1998; Goussard et al. 1998; Strobel et al 1999; Strobel & Weigel 2001). Nevertheless, it must be remarked that for a value of gravitational mass equal to 1.4 $M_\odot$ (the canonical mass for compact star candidates), the difference between the predicted radii of nonrotating configurations of SS and NS amounts, at most, only to about 5 km; a value that cannot be directly observed. There arises, therefore, a necessity to heavily rely on models of astrophysical phenomena associated with systems containing a compact star to estimate the radius: for isolated pulsars, models of glitches (e.g. Datta & Alpar 1993; Link et al. 1992) have been used in the past for making estimates of the structure parameters and for compact stars in binaries, such estimates have been made by appropriately modelling photospheric expansion in X–ray bursts (van Paradijs 1979; Goldman 1979) and more recently by constraining the inner–edge of accretion discs and demanding that the radius of the compact star be located inside this inner–edge (Li et al. 1999a; Li et al. 1999b; Burderi & King 1998; Psaltis & Chakrabarty 1999). In particular, the work by Li et al. (1999a; 1999b) suggest strange stars as possible accretors. However, these calculations did not include the full effect of general relativity. Even on inclusion of these effects (Bombaci et al 2000), the results for at least one source: 4U 1728-34, remain unchanged. There have also been contradictory reports on the existence of strange stars: for example, calculations of magnetic field evolution of SSs over dynamical timescales, make it difficult to explain the observed magnetic field strengths of isolated pulsars (Konar 2000). On the other hand, Xu & Busse (2001) show that SSs may possess magnetic fields, having the observed strengths. These magnetic fields, these authors argue, originate due to dynamo effects. In our analysis here, we ignore the effects of magnetic field.

---

[1] In the present work, we consider only *bare* strange stars, *i.e.* we neglect the possible presence of a crust of normal (confined) matter above the deconfined quark matter core (see *e.g.* Alcock, Farhi & Olinto 1986).

In this paper, we calculate constant gravitational mass equilibrium sequences of rotating SSs, considering the full effect of general relativity. We solve Einstein field equations and the equation for hydrostatic equilibrium simultaneously for different SS equations of state (EOS) models. We compare our theoretical results with those obtained previously for NSs (Bhattacharyya et al. 2000). In addition, we calculate the radial profiles of effective temperature in accretion discs around SSs. These profiles are important inputs in accretion disc spectrum calculations, crucially depending on the radius of the inner edge of the accretion disc. This radius is determined by the location of $r_{\mathrm{orb}}$ with respect to that of the surface (R) of the star, both of which are sensitive to the EOS, through the rotation of the central object. In particular, we notice that $r_{\mathrm{orb}}$ increases with stellar angular momentum ($J$) beyond a certain critical value (a property not seen in either rotating black holes or neutron stars). We trace this behaviour to the dependence of $dr_{\mathrm{orb}}/dJ$ on the rate of change of the radial gradient of the Keplerian angular velocity at $r_{\mathrm{orb}}$ with respect to $J$. The prospect of using the temperature profiles for calculation of accretion disc spectrum and subsequent comparison with observational data, therefore, gives rise to the possibility of constraining SS EOS, and eventually to distinguish between SSs and NSs.

The structure of the paper is as follows. In section 2, we describe briefly our method of calculation of the structure of rapidly rotating relativistic stars and the accretion disc temperature profile. We also provide here, a brief description of the EOS models used. The results described in section 3 are discussed and summarised in section 4.

## 2. Methodology

### 2.1. Structure calculation

We calculate the structure of rapidly rotating relativistic stars using an iterative procedure described in Cook et al. (1994) (see also Datta et al. 2000). The metric describing the space–time around a rotating relativistic star can be given as

$$
\begin{aligned}
dS^2 \; = \; & g_{\mu\nu}dx^\mu dx^\nu (\mu, \nu = 0, 1, 2, 3) \\
= \; & -e^{\gamma+\rho}dt^2 + e^{2\alpha}(d\bar{r}^2 + \bar{r}^2 d\theta^2) \\
& + e^{\gamma-\rho}\bar{r}^2 \sin^2\theta (d\phi - \omega dt)^2
\end{aligned}
\tag{1}
$$

This metric is stationary, axisymmetric, asymptotically flat and reflection-symmetric (about the equatorial plane). The metric potentials $\gamma, \rho, \alpha$ and the angular speed ($\omega$) of zero-angular-momentum-observer (ZAMO) with respect to infinity, are all functions of the quasi-isotropic radial coordinate ($\bar{r}$) and polar angle ($\theta$). The quantity $\bar{r}$ and the Schwarzschild–like radial coordinate ($r$) are related through the coordinate transformation: $r = \bar{r}e^{(\gamma-\rho)/2}$. We use the geometric units $c = G = 1$ in all the equations that appear this paper.

With the assumption that the star is rigidly rotating and described by a perfect fluid, we solve Einstein field equations and the equation of hydrostatic equilibrium self-consistently and numerically from the centre of the star upto infinity to obtain $\gamma, \rho, \alpha, \omega$ and $\Omega_*$ (angular speed of neutron star with respect to an observer at infinity) as functions of $\bar{r}$ and $\theta$. The inputs for this calculation are: the chosen EOS, assumed values of central density and ratio of polar to equatorial radii. The outputs are bulk structure parameters such as: gravitational mass ($M$), equatorial radius ($R$), angular momentum ($J$), moment of inertia ($I$) etc. of the compact star. We also calculate the specific disc luminosity ($E_{\mathrm{D}}$), the specific boundary layer luminosity ($E_{\mathrm{BL}}$), the radius ($r_{\mathrm{orb}}$) of the innermost stable circular orbit (ISCO), specific energy ($\tilde{E}$), specific angular momentum ($\tilde{l}$) and angular speed ($\Omega_{\mathrm{K}}$) of a test particle in Keplerian orbits (Thampan & Datta 1998).

In the present work, for each adopted EOS, we construct constant $M$ equilibrium sequences with $\Omega_*$ varying from the non-rotating case (static limit; $\Omega_* = 0$) upto the centrifugal mass shed limit (rotation rate at which inwardly directed gravitational forces are balanced by outwardly directed centrifugal forces; $\Omega_* = \Omega$ms) (Bhattacharyya et al. 2000).

| EOS label | compact star | EOS model |
|:---:|:---:|:---:|
| A | SS | Dey et al. (1998),  their model SS1 |
| B | SS | Farhi and Jaffe (1984),  $B = 90$ MeV/fm$^3$, $m_s = 0$ |
| C | SS | Farhi and Jaffe (1984),  $B = 60$ MeV/fm$^3$, $m_s = 200$ MeV |
| D | SS | Farhi and Jaffe (1984),  $B = 60$ MeV/fm$^3$, $m_s = 0$ |
| E | NS | Pandharipande (1971),  hyperonic matter |
| F | NS | Baldo et al. (1997),  nuclear matter |
| G | NS | Sahu et al. (1993),  nuclear matter |

**Table 1.** The list of EOS models used in this work.

## 2.2. Disc temperature profile calculation

The effective temperature of a geometrically thin blackbody disc is given by

$$T_{\text{eff}}(r) = (F(r)/\sigma)^{1/4} \qquad (2)$$

where $\sigma$ is the Stefan-Boltzmann constant and $F$, the X-ray energy flux (due to viscous dissipation) per unit surface area. We calculate $F$ using the expression (Page & Thorne 1974; valid for a geometrically thin non–self–gravitating disc embedded in a general axisymmetric space–time of a rotating black hole):

$$F(r) = \frac{\dot{M}}{4\pi r} f(r) \qquad (3)$$

where

$$f(r) = -\Omega_{\text{K},r}(\tilde{E} - \Omega_{\text{K}}\tilde{l})^{-2} \int_{r_{\text{in}}}^{r} (\tilde{E} - \Omega_{\text{K}}\tilde{l})\tilde{l}_{,r} dr \qquad (4)$$

Here $r_{\text{in}}$ is the disc inner edge radius and a comma followed by a variable as subscript to a quantity, represents a derivative of the quantity with respect to the variable.

Although formulated for the case of black holes, these expressions also hold for NS (Bhattacharyya et al. 2000). Depending on the mass of a compact star and the EOS describing it, there exist two possibilities for the location of ISCO *vis-á-vis* the surface of the star: $R > r_{\text{orb}}$, or $R < r_{\text{orb}}$. We assume that if $R > r_{\text{orb}}$, the inner–edge of the accretion disc touches the star and we take $r_{\text{in}} = R$; otherwise, $r_{\text{in}} = r_{\text{orb}}$. For strange stars, the inner–edge of the disc rarely touches the surface of the star (as described in the results section). So the above expressions will almost always be exactly valid for SSs. Using these results, it is straightforward to calculate the temperature profile of the accretion disc as a function of $M$ and $\Omega_*$ of the central star for any chosen EOS.

## 2.3. Equation of state

For strange quark matter we use two phenomenological models for the EOS. The first one is a simple EOS (Farhi and Jaffe 1984) based on the MIT bag model for hadrons. We begin with the case of massless, non-interacting (*i.e.* QCD structure constant $\alpha_{\text{c}} = 0$) quarks and with a bag constant $B = 60$ MeV/fm$^3$ (hereafter EOS D). Next, we consider a finite value for the mass of the strange quark within the same MIT bag model EOS. We take $m_{\text{s}} = 200$ MeV and $m_{\text{u}} = m_{\text{d}} = 0$, $B = 60$ MeV/fm$^3$, and $\alpha_{\text{c}} = 0$ (EOS C). To investigate the effect of the bag constant, we take (almost) the largest possible value of $B$ for which SQM is still the ground state of strongly interacting matter, according to the strange matter hypothesis. For massless non-interacting quarks this gives $B = 90$ MeV/fm$^3$ (EOS B). The second model for SQM is the EOS given by Dey *et al.* (1998), which is based on a different quark model than the MIT bag model. This EOS has asymptotic freedom built in, shows confinement at zero baryon density, deconfinement at high density, and, for an appropriate choice of the EOS parameters entering the model, gives absolutely stable SQM according to the strange matter hypothesis. In the model by Dey *et al.* (1998), the quark interaction is described by a screened inter–quark vector potential originating from gluon exchange, and by a density-dependent scalar potential which restores chiral symmetry at high density (in the limit of massless quarks). The density-dependent scalar potential arises from the density dependence of the in-medium effective quark masses $M_q$, which are taken to depend upon the baryon number density $n_B$ according to $M_q = m_q + 310\mathrm{MeV} \times sech\left(\nu\frac{n_B}{n_0}\right)$, where $n_0$ is the normal nuclear matter density, q($=$ u, d, s) is the flavor index, and $\nu$ is a parameter. The effective quark mass $M_q(n_B)$ goes from its constituent masses at zero density, to its current mass $m_q$, as $n_B$ goes to infinity. Here we consider a parameterization of the EOS by Dey *et al.* (1998), which corresponds to the choice $\nu = 0.333$ for the parameter entering in the effective quark mass, and we denote this model as EOS A.

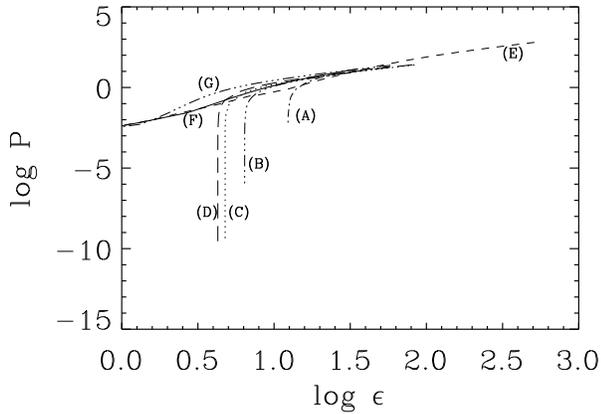

**Fig. 1.** Logarithmic plot of pressure vs. matter density for the EOS models used here. The density and pressure are in units of $1.0 \times 10^{14}$ g cm$^{-3}$ and $(1.0 \times 10^{14})$ $c^2$ cgs respectively.

For NSs, we use three representative equations of state which span a wide range of *stiffness*. A very soft EOS is the one for hyperonic matter given by Pandharipande (1971) which we denote as EOS E. As a representative stiff equation of state, we use the EOS by Sahu, Basu & Datta (1993) (hereafter EOS G). Finally, we consider the EOS by Baldo, Bombaci & Burgio (1997), which is a microscopic EOS for $\beta$-stable nuclear matter based on the Argonne $v14$ nucleon-nucleon interaction implemented by nuclear three-body forces (EOS F). The latter EOS is intermediate in stiffness with respect to the previous two EOS models.

A list of the designation along with the salient features of the EOS models used here is provided in Table 1.

We also display the qualitative variations in these EOS models in a log–log plot of Fig. 1. The differences between SS and NS EOS are plainly evident, especially at lower pressures.

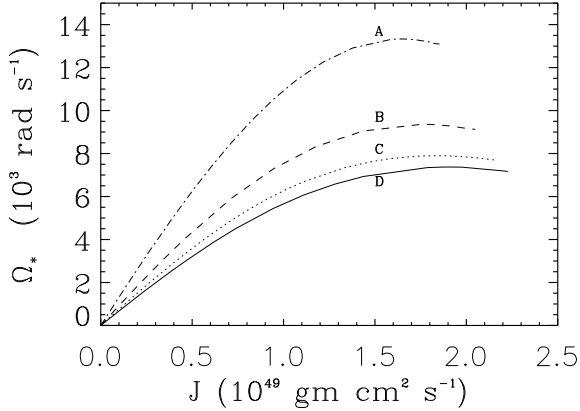

**Fig. 2.** Angular speed ($\Omega_*$) as a function of total angular momentum ($J$) for strange star. The curves are labelled by the nomenclature of Table 1 and are for a fixed gravitational mass ($M = 1.4\ M_\odot$) of the strange star.

## 3. The Results

We have calculated the structure parameters and the disc temperature profiles for rapidly rotating, constant gravitational mass sequences of SSs in general relativity. The results for SS are compared with those for NS. For illustrative purposes here, we have chosen the value of gravitational mass to be $1.4\ M_\odot$.

Fig. 2, depicts the variation of $\Omega_*$ with the total angular momentum ($J$) for constant gravitational mass and for the four SS EOS. The curves extend from the static limit to the mass-shed limit. The striking feature here is that, although $J$ increases monotonically from slow rotation to mass-shed limit, $\Omega_*$ shows a non-monotonic behaviour: maximum value of $\Omega_*$ (i.e. $\Omega_*^{\mathrm{max}}$) occurs at a value of $J$ lower than that for mass–shed limit. Although this seems to be a generic feature for SS EOS, $\Omega_*$ is always a monotonic function of $J$ for constant gravitational mass NS sequences and hence constitutes an essential difference between SS and NS (see section 4 for discussions). Our calculations show that at maximum $\Omega_*$, the ratio of rotational kinetic energy to total gravitational energy: $T/W$ approaches the value of 0.2 (see next paragraph). It has been pointed out by Gourgoulhon et al., 1999 that such high values of $T/W$ make the configurations unstable to triaxial instability. It can also be noticed that for stiffer EOS, the star possesses a higher value of $J$ at mass shed limit ($\Omega_*^{\mathrm{max}}$ also occurs proportionately at higher $J$).

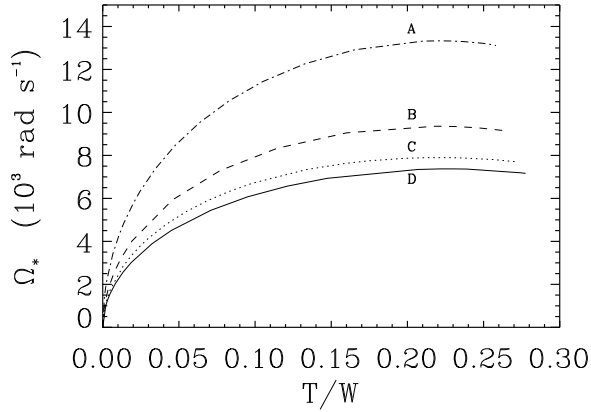

**Fig. 3.** Angular speed ($\Omega_*$) as a function of the ratio of rotational kinetic energy and gravitational binding energy ($T/W$) for strange star. Curve labels have the same meaning as in Fig. 2.

In order to expand upon the results of Fig. 2, we plot $\Omega_*$ vs. $T/W$ for various SS EOS in Fig. 3. It is seen that for all SS EOS, $T/W$ becomes greater than 0.25 at mass-shed limit, while for NS EOS it is usually between 0.1 and 0.14 (Cook et al., 1994). Interestingly, for all SS EOS, $\Omega_*^{\max}$ occurs at about the same value of $T/W$ ($\approx 0.2$).

Fig. 4 displays $\Omega_*$ and $\Omega_{\mathrm{in}}$ (i.e. the Keplerian angular speed of a test particle at $r_{\mathrm{in}}$) against $J$. The four panels are for the four SS EOS we use. We notice the interesting behaviour that $\Omega_*$ and $\Omega_{\mathrm{in}}$ curves cross each other at a point near $\Omega_*^{\max}$. For rotating NS configurations, since the equality $r_{\mathrm{in}} = R$ is almost always (except for very soft EOS models: Fig. 1, Bhattacharyya et al. 2000) achieved for rotation rates well below that at mass-shed limit (for $M = 1.4\ M_\odot$), always $\Omega_* \leq \Omega_{\mathrm{in}}$ (the equality is achieved only at mass shed limit). On the other hand, for SS, $r_{\mathrm{orb}}$ is almost always greater than $R$ (as explained in the next paragraph) and when the star approaches Keplerian angular speed at the equator, $\Omega_*$ becomes greater than $\Omega_{\mathrm{in}}$.

Fig. 5 is a plot of the variation of $r_{\mathrm{in}}$ and $R$ with $\Omega_*$ for four SS EOS. We see that the behaviour of $R$ is monotonic from slow rotation to the mass-shed limit, even though that of $\Omega_*$ is not. As mentioned earlier for all $\Omega_*$ from static limit upto mass shed limit, $r_{\mathrm{in}} > R$ for 3 SS EOS. Only for the stiffest SS EOS, that we have chosen, does the disc touch the star (for an intermediate value of $\Omega_*$). This is distinct from the case of NS (see Fig. 1 of Bhattacharyya et al. 2000). The reason for such a behaviour is the non–monotonic variation of $r_{\mathrm{orb}}$ with $J$ for SS (contrary to the case of NS and black holes); this is discussed further in the next section.

In Fig. 6, we plot the variation of $r_{\mathrm{in}}$ with $\Omega_*$ for three SS EOS and two NS EOS: for each case, our softest EOS and our stiffest EOS have been chosen. In addition, we display the corresponding results for EOS model B too. It is clear that in the $r_{in}$–$\Omega_*$ space, there exists a region that is spanned by both NS and SS configurations. Interestingly, however, there also exists certain regions occupied exclusively by either SS or NS configurations. The possible observational consequences of this result is discussed in the next section.

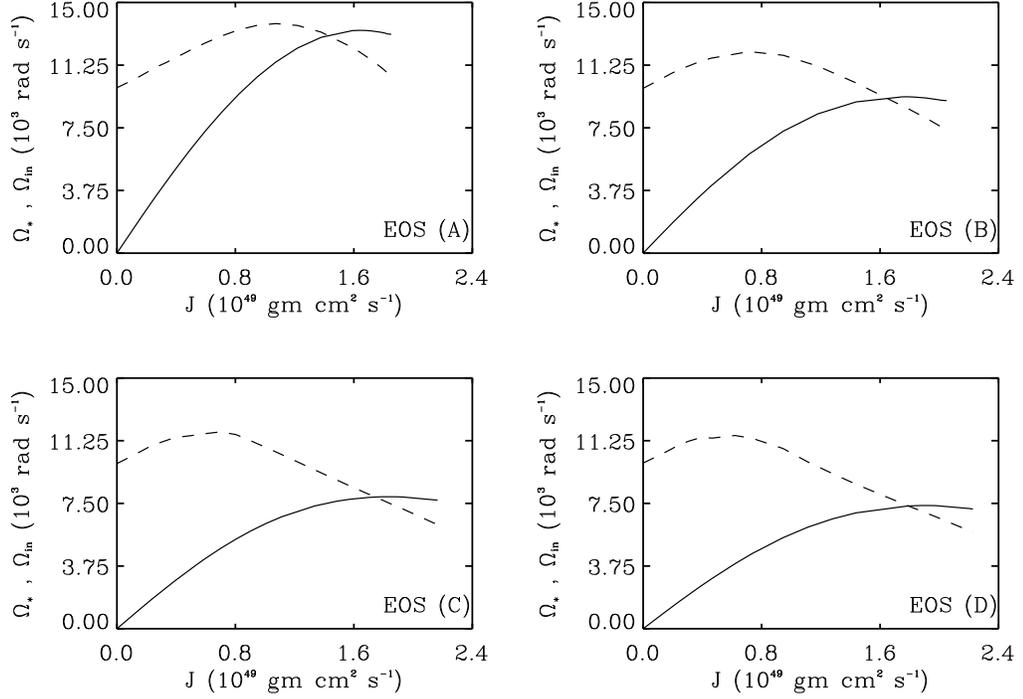

**Fig. 4.** Angular speed ($\Omega_*$) of the strange star (solid curve) and the Keplerian angular speed ($\Omega_{in}$) of a test particle at the inner edge of the disc (dashed curve) as functions of total angular momentum ($J$) of strange star. The curves are for a fixed gravitational mass ($M = 1.4\ M_\odot$) of the strange star. Different panels are for different SS EOS models.

Fig. 7 displays the radial profiles of temperature:(i) assuming a purely Newtonian accretion disc and (ii) considering general relativistic accretion discs for (a) SS (EOS D) and (b) NS (EOS F), each represented by two configurations: the non–rotating and mass shed for $M = 1.4\ M_\odot$. We also display the temperature profile (curve 5) for a SS configuration of $M = 1\ M_\odot$, described by EOS (A) (the constraints obtained by Li et al. 1999a; 1999b) and having a period $P = 2.75$ ms (the mass and period corresponding to that inferred for the source 4U 1728-34: Méndez & Van der Klis 1999). It must be remembered that in this figure (and the next), curve 5 represents the temperature profile for a different $M$ value than the rest of the curves and is displayed in the same figure, only for illustrative purposes. From this figure we see that for $M = 1.4\ M_\odot$, the Newtonian value of temperature is about 25% higher than the general relativistic value near the inner edge of the disc. This shows the importance of general relativity and rotation near the surface of the star. The difference between the effects of SS EOS and NS EOS on temperature profiles (at the inner portion of the disc) is also prominent at mass-shed limit (due to the difference in rotation rates for these two configurations). Such differences in temperature profiles are also expected to show up in the calculations of spectra at higher energies.

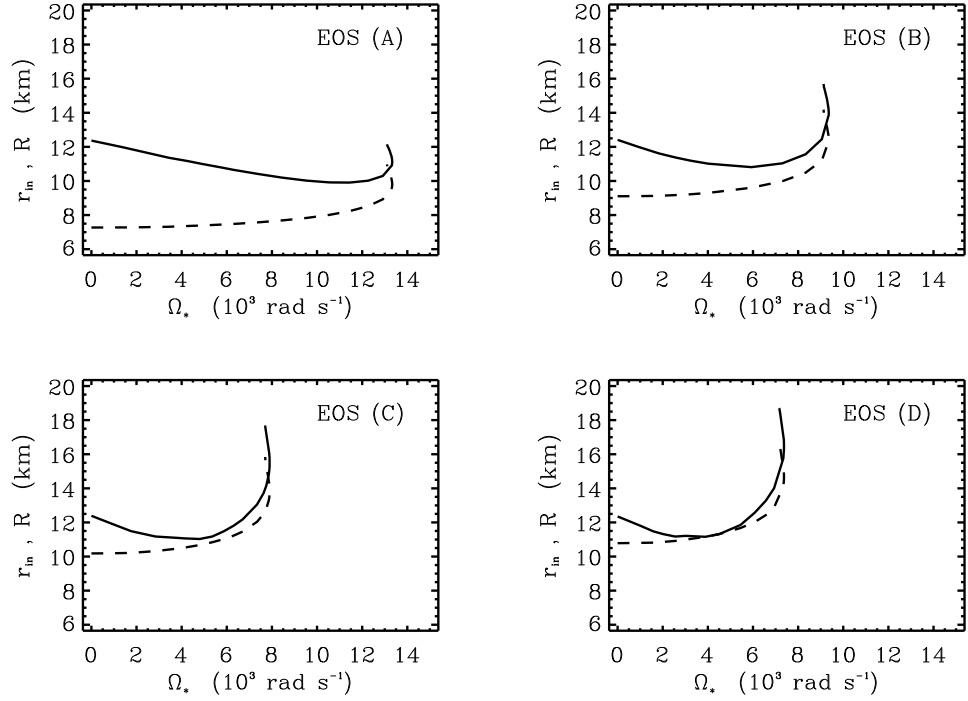

**Fig. 5.** Disk inner edge radius ($r_{\rm in}$, solid curve) and strange star radius ($R$, dashed curve), as functions of angular speed $\Omega_*$ for various EOS models. The curves are for a fixed gravitational mass ($M = 1.4\ M_\odot$) of the strange star.

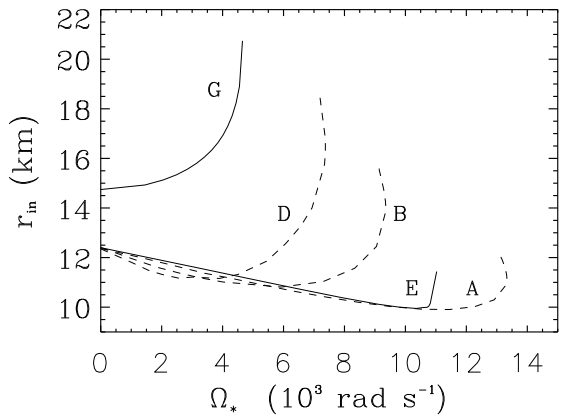

**Fig. 6.** Disk inner edge radius ($r_{\rm in}$) as a function of angular speed $\Omega_*$ of the compact star. The curves have their usual meaning.

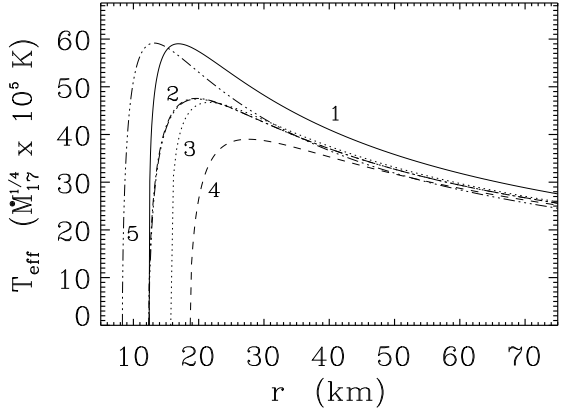

**Fig. 7.** Accretion disc temperature profiles: Curve (1) corresponds to the Newtonian case, curve (2) to the Schwarzschild case (coincident curves for NS EOS model F and SS EOS model D), curve (3) to a neutron star (EOS model F) rotating at the centrifugal mass-shed limit and curve (4) to a strange star (EOS model D) rotating at the centrifugal mass-shed limit. For curve (1) it is assumed that, $r_{in} = 6GM/c^2$. The curves (1–4) are for a fixed gravitational mass ($M = 1.4 \ M_\odot$) of the compact star. Curve (5) corresponds to a configuration that has $M = 1 \ M_\odot$ and $\Omega_*$ corresponding to a period $P = 2.75$ ms (inferred for 4U 1728-34; see text) and described by EOS model A. In this and all subsequent figures, the temperature is expressed in units of $\dot{M}_{17}^{1/4} \ 10^5$ K, where $\dot{M}_{17}$ is the steady state mass accretion rate in units of $10^{17}$ g s$^{-1}$.

In the panel (a) of Fig. 8, we display the temperature profiles for configurations (as in Fig. 7) composed of SS EOS (D) (curves 1–4), represented by different $\Omega_*$ (corresponding to $\Omega_* = 0$, minimum $r_{in}$, $\Omega_* = \Omega_*^{max}$ and mass-shed limit); curve (5) is the same as in Fig. 7. The behaviour of temperature profiles is non-monotonic with $\Omega_*$. The panel (b) shows the temperature profiles at mass-shed for various SS EOS along with curve (5). Here the temperature profiles show monotonic behaviour with the stiffness of EOS. The behaviour of the temperature profiles in both the panels are similar to those calculated for NS (Bhattacharyya et al. 2000). Notice the substantial difference in the maximum temperature; sufficiently sensitive observations are, therefore, expected to complement the findings of Li et al. (1999a; 1999b).

The variations of $E_D$, $E_{BL}$, the ratio $E_{BL}/E_D$ and $T_{eff}^{max}$ with $\Omega_*$ are displayed in Fig. 9. Each plot contains curves corresponding to all the SS EOS models considered here. The behaviour of all the curves are similar to those for any NS EOS (see Fig. 5 of Bhattacharyya et al. 2000). The only difference being that due to the non-monotonic behaviour of $\Omega_*$ from slow rotation to mass-shed limit for SS EOS, making the curves turn inward at the terminal (mass shed) rotation rate.

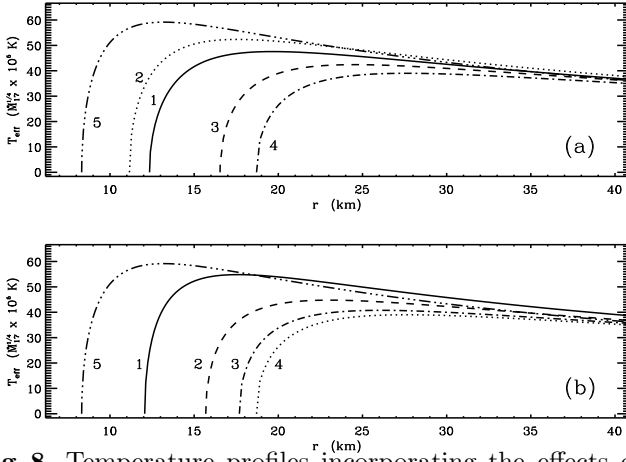

**Fig. 8.** Temperature profiles incorporating the effects of rotation of the strange star. The plots correspond to (a) EOS model D and an assumed strange star mass of $M = 1.4\ M_\odot$ (curves 1–4) for rotation rates: $\Omega_* = 0$ (curve 1), $\Omega_* = 3.891 \times 10^3$ rad s$^{-1}$ (curve 2), $\Omega_* = 7.373 \times 10^3$ rad s$^{-1}$ (curve 3), $\Omega_* = 7.163 \times 10^3$ rad s$^{-1}$ = $\Omega_{ms}$ (curve 4), (b) the same assumed mass and $\Omega_* = \Omega_{ms}$ for the four EOS models (A):curve 1, (B):curve 2, (C):curve 3 and (D):curve 4. In both panels, curve (5) is the same as that in Fig. 7.

In Fig. 10, we make a comparison between SS and NS for the same quantities displayed in Fig. 9. We have used three SS EOS and two NS EOS models (the softest and the stiffest for each case). In all the panels, SS and NS both are seen to have their own exclusive regions in the high and low $\Omega_*$ parameter space respectively. This is especially prominent for $E_{BL}$ and $E_{BL}/E_D$. We also notice that for SS, at $\Omega_* = \Omega_*^{max}$, the values of $E_{BL} \approx 0.05$ and $E_{BL}/E_D \approx 1.0$ for all EOS. On the contrary, for neutron stars, both $E_{BL}$ and $E_{BL}/E_D$ become $\approx 0$ at $\Omega_*^{max}$ $(= \Omega_{ms})$.

## 4. Summary and Discussion

In this paper we have calculated the structure parameters and the disc temperature profiles for rapidly rotating SSs (for constant gravitational mass sequence with $M = 1.4\ M_\odot$ ) and compared them with those for NSs with the aim of finding possible ways to distinguish between the two. For the sake of completeness, we have compared the properties of these two types of stars all the way from slow rotation to mass-shed limit.

The striking feature of SSs is the non–monotonic behaviour of $\Omega_*$ with $J$ such that $\Omega_*^{max}$ occurs at lower value of $J$ than that of the mass–shed limit. Hence the other SS structure parameters become non-monotonic functions of $\Omega_*$. This behaviour is observed even for the constant rest mass sequences of SS (e.g. Gourgoulhon et al 1999; Bombaci et al. 2000). In contrast, for NSs, the structure parameters are all monotonic functions of $\Omega_*$. An implication of the non–monotonic behaviour of $\Omega_*$ with $J$ is that if an isolated sub–millisecond pulsar is observed to be spinning up, it is likely to be a SS rather than a NS.

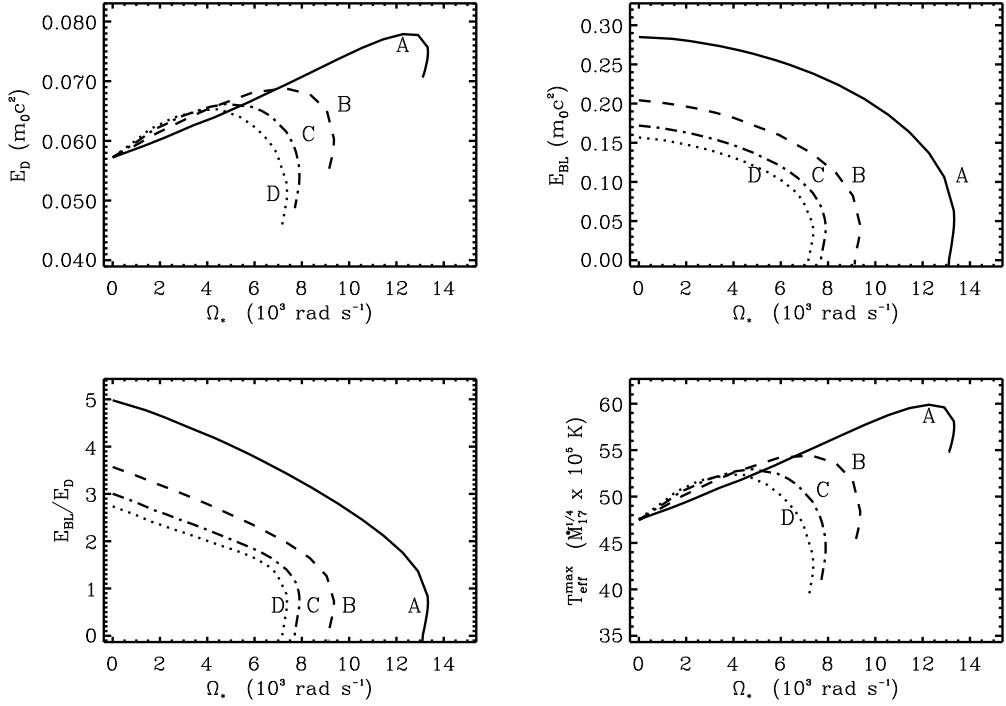

**Fig. 9.** The variations of the $E_\mathrm{D}$, $E_\mathrm{BL}$, $E_\mathrm{BL}/E_\mathrm{D}$ and $T_\mathrm{eff}^\mathrm{max}$ with $\Omega_*$ for a chosen strange star mass value of 1.4 $M_\odot$ for the four SS EOS models. The curves have the same significance as Fig. 3.

Because of higher values of $T/W$ ($\gtrsim 0.2$), SSs are more prone to secular instabilities compared to NSs at rapid rotation (Gourgoulhon et al. 1999). Our calculations show that at $\Omega_*^\mathrm{max}$, $T/W > 0.2$.

Another important feature of SS gravitational mass sequence (in contrast to the corresponding NS sequences) is the crossing point in $\Omega_*$ and $\Omega_\mathrm{in}$. This feature has important implication in models of kHz QPOs: for example if $\Omega_*$ is greater than $\Omega_\mathrm{in}$, the beat–frequency models ascribing higher frequency to Keplerian frequencies will not be viable.

It can be noted from Fig. 4, that with the increase in stiffness of the EOS models, $J_\mathrm{cross}$ increases and $\Omega_\mathrm{*,cross}$ (the subscript "cross" corresponds to the point $\Omega_\mathrm{in} = \Omega_*$) decreases monotonically. It is also seen that in general all the quantities vary monotonically with the stiffness for both SS and NS EOS (see also Bhattacharyya et al. 2000).

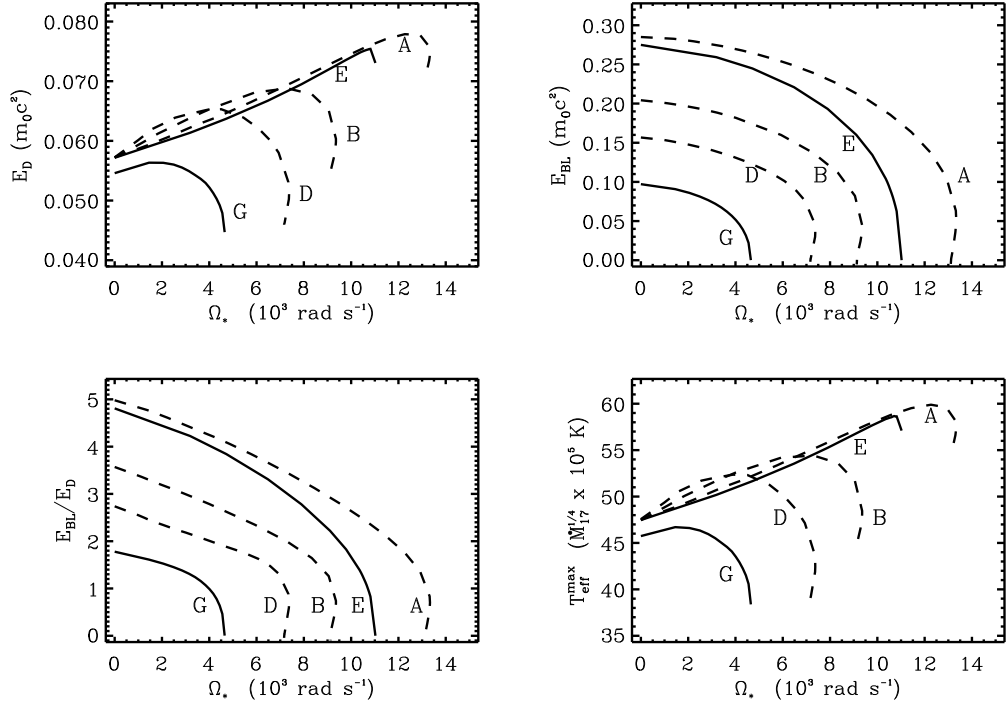

**Fig. 10.** Same as Fig. 9, except the fact that here two NS EOS models and two SS EOS models are used. The curves have the same significance as Fig. 6.

For SSs, the inner–edge of the accretion disc rarely touches the surface of the star (even for maximum rotation rates), while for rapidly rotating NSs, the accretion disc extends upto the stellar surface for almost all rotation rates. Since the inner accretion disc boundary condition is different for both these cases, we expect important observable differences (both temporal and spectral) in X–ray emission (from the boundary layer and the inner accretion disc) from SSs and NSs.

A brief note on the variation of $r_{\rm orb}$, with specific angular momentum is in order here. As mentioned earlier, beyond a certain value of the angular momentum, the radius of the ISCO increases with increasing angular momentum - a property not seen either in the case of NSs or black holes. The reason for this can be traced to the radial gradient of the angular velocity of the particles at the marginally stable orbit and the analysis is described as follows:

For the metric described by equation (1), the second derivative of the effective potential may be written as (Cook et al 1994; Thampan & Datta 1998)

$$
\begin{aligned}
\tilde{V}_{,rr} \equiv\ & 2\left[\frac{r}{4}(\lambda_{,r}^2 - \gamma_{,r}^2) - \frac{1}{2}e^{-2\lambda}\omega_{,r}^2 r^3 - \lambda_{,r} + \frac{1}{r}\right]\tilde{v}^2 \\
& + [2 + r(\gamma_{,r} - \lambda_{,r})]\tilde{v}\tilde{v}_{,r} - e^{-\lambda}\omega_{,r}r\tilde{v} \\
& + \frac{r}{2}(\gamma_{,r}^2 - \lambda_{,r}^2) - e^{-\lambda}r^2\omega_{,r}\tilde{v}_{,r} = 0
\end{aligned}
\tag{5}
$$

Simplification of this, using the other equations of motion (Bardeen 1970), yield

$$r^2(1-v^2)\tilde{V}_{,rr} = -X\left[\frac{r\Omega_{,r}}{\Omega-\omega} + \frac{1-v^2}{2v^2}X\right] \quad (6)$$

Where $X = v^2(2+r\gamma_{,r}-r\rho_{,r})+r(\gamma_{,r}+\rho_{,r})$. The marginal stability criterion, therefore, yields the rate of change of the marginal stable orbit, with respect to specific angular momentum ($j = J/M^2$) as:

$$r_{\text{orb}\,,j} = r_{\text{orb}}\left\{\frac{(\Omega-\omega)_{,j}}{\Omega-\omega} - \frac{\Omega_{,rj}}{\Omega_{,r}} + 2\frac{v_{,j}}{v(v^2-1)} + \frac{X_{,j}}{X}\right\} \quad (7)$$

where the terms in the parenthesis are to be evaluated at $r_{\text{orb}}$. We calculate the four terms in the parenthesis in equation (7) and find that the second term dominates the net rate of change of $r_{\text{orb}}$ with $j$. Which implies that at the value of $j$ where $r_{\text{orb},j}$ changes sign, although the first three terms are observed to change sign, the net sign is only dependent on that of $\Omega_{,rj}$ at ISCO.

From Fig. 6 we see that for $\Omega_*$ in the range (0,4028) rad. s$^{-1}$ (the second quantity in the range is the rotation rate of PSR 1937+21: Backer et al. 1982, the fastest rotating pulsar observed so far), a major portion of the $r_{\text{in}}$-$\Omega_*$ space is occupied exclusively by NS. So if $r_{\text{in}}$ can be determined independently from observations (for example, by fitting the soft component of the observed spectrum by the XSPEC model "discbb" available in XANADU: see for example Kubota et al 1998, or, from the observed kHz QPO frequencies), there is a fair chance of inferring the central accretor to be a NS rather than SS (provided the mass of the central accretor is known by other means). This is also applicable to $E_{\text{BL}}$ and $E_{\text{BL}}/E_{\text{D}}$ (Fig. 10). It is also to be noted that Li et al. (1999 a; 1999b) did a similar search in the $M-R$ parameter space and concluded the millisecond X–ray pulsar SAX J1808.4-3658 and the central accretor in 4U 1728-34 to be likely SSs. If, indeed this is true, then it is possible to constrain the *stiffness* of the equation of state of SQM (Bombaci 2000), and to exclude EOS models (like EOS C and EOS D) stiffer than our EOS B.

Calculation of the accretion disc spectrum involves the temperature profiles as inputs. The spectra of accretion discs, incorporating the full effects of general relativity for NSs (Bhattacharyya et al. 2000;2001a;2001b) show sensitive dependence on the EOS of high density matter. The similarity in the values of the maximum disc temperature implies a similar indistinguishability in the spectra of SS as compared to NS. Nevertheless, just as $E_{\text{BL}}$ and other quantities show that NS exclusively occupy certain regions in the relevant parameter space, we expect that it will be possible to make a differentiation between these two compact objects by modelling the boundary layer emission. If as mentioned in previous paragraph, we exclude EOS models stiffer than B, then from Fig. 10, we see that a fairly accurate measurement of $E_{\text{BL}}$ ($L_{\text{BL}}/\dot{M}c^2$) and $\Omega_*$ can indicate whether the central accretor is a NS or SS if the corresponding point falls outside the strip defined by curves B and E.

The current uncertainties in theoretical models of boundary layer emission and the variety of cases presented by models of rotating compact objects, calls into order, a detailed investigation into these aspects of LMXBs - especially with the launch of new generation X-ray satellites (having better sensitivities and larger collecting areas) on the anvil.

*Acknowledgements.* SB thanks the Director, Raman Research Institute, Bangalore for facilities provided, Director, IUCAA, Pune for kind hospitality and Pijush Bhattacharjee for encouragement and discussions.